\documentclass[12pt]{article}
\usepackage[latin1]{inputenc}
\usepackage{pstricks}
\usepackage{amsmath}
\usepackage{float}
\usepackage{color}
\input epsf
\usepackage{amssymb}
\usepackage[dvips]{graphicx,psfrag}
\newcommand{\be}{\begin{equation}}
\newcommand{\ee}{\end{equation}}
\newcommand{\bea}{\begin{eqnarray}}
\newcommand{\eea}{\end{eqnarray}}
\newcommand{\lb}{\label}
\begin{document}
\begin{titlepage}
\title{The complementarity of the redshift drift}
\author{B. Moraes$^{a,b}$\thanks{email:bruno.moraes@cbpf.br}~and
D. Polarski$^c$\thanks{email:david.polarski@univ-montp2.fr}\\
\hfill\\
$^a$~Centro Brasileiro de Pesquisas F\'isicas - CBPF, Brazil\\
$^b$~Laborat\'{o}rio Interinstitucional de e-Astronomia - LIneA, Brazil\\
$^c$~Lab. Charles Coulomb, CNRS, Universit\'e Montpellier 2, France}
\pagestyle{plain}
\pagestyle{plain}
\date{\today}

\pagestyle{plain}
\date{\today}

\maketitle
\thispagestyle{empty}
\begin{abstract}
We derive some basic equations related to the redshift drift and we 
show how some dark energy (DE) properties can be retrieved from it. 
We consider in particular three kinds of DE models which exhibit a 
characteristic signature in their redshift drift while no such signature 
would be present in their luminosity-distances: a sudden change of the 
equation of state parameter $w_{DE}$ at low redshifts, oscillating DE 
and finally an equation of state with spikes at low redshifts. Accurate 
redshift drift measurements would provide interesting complementary 
probes for some of these models and for models with varying gravitational 
coupling. While the redshift drift would efficiently constrain models with 
a spike at $z\sim 1$, the signature of the redshift drift for models with 
large variations at very low redshifts $z<0.1$ would be unobservable, 
allowing a large arbitrariness in the present expansion of the universe.
\end{abstract}

\end{titlepage}


\section{Introduction}
Data suggests that our universe has entered a stage of accelerated expansion
rate. 
This is a radical departure from conventional cosmology in which the expansion
was 
constantly decelerated except for the inflationary stage in the very early
universe. 
The mechanism behind this late-time accelerated expansion is unclear and many
models 
were suggested \cite{SS00}. 
If it is caused by some isotropic perfect fluid component called dark energy
(DE) 
while gravity is governed by General Relativity (GR), then DE should make up
about 
two thirds of the universe content.  
The simplest possibility, a cosmological constant $\Lambda$ is problematic from
the 
theoretical point of view because of its tiny amplitude.
Further while the data is in agreement with a cosmological constant $\Lambda$,
the 
precision of these data do not allow to rule out models where the equation of
state 
(EoS) parameter of the dark energy component $w_{DE}$ differs from $-1$ and
varies 
with time, and in which the corresponding energy density depends on time as
well. 
A great number of models have been proposed and all their properties were 
intensively investigated in the hope that future accurate observations would
allow 
to discriminate between these models and that only a small part of them would
remain 
as viable candidates while the other models would be ruled out.
Another attractive possibility is a modification of gravity on cosmic scales. 
Virtually all the foundations of standard cosmology have been questioned in the 
quest for a solution to the dark energy problem. 
Decisive progress in settling the issue will come from observations. 
Experiments of different kinds are planned that will probe with exquisite
precision 
the universe background expansion and also the evolution of the perturbations. 

To make progress in the understanding of the nature of DE it is desirable to
explore all 
ways in which its properties can be probed. One such probe is the redshift
drift. The 
feasibility of such measurements was reassessed in \cite{L98}, many years after
its 
theoretical discovery \cite{S62}.
It has been suggested recently as a way to test general properties of the DE
paradigm 
\cite{BQ07}. These measurements could give additional insight into the
properties 
of DE because they probe directly the quantity $H(z)$.
It is this aspect that we want to emphasize and to investigate in this work. 
Before embarking on the quantitative assessment of redshift drift data 
for some specific models we first derive some general properties of the 
redshift drift relevant for the study of DE. 

\section{The redshift drift}
In an expanding universe many physical quantities evolve with time. The redshift
suffered 
by radiation emitted by a source depends on time as well. Indeed let us consider
radiation 
emitted by a source at the emission time $t_e$. If this radiation is observed at
time $t_0$, 
the corresponding redshift is given by 
\be
1 + z(t_e,t_0) = \frac{a(t_0)}{a(t_e)}\equiv \frac{a_0}{a_e}~,
\ee
where we use the notation $z(t_e,t_0)$ to emphasize that the redshift depends on
the 
emission time $t_e$ and on the observation time $t_0$. If radiation emitted by
the same 
source is observed at time $t_0 + \delta t_0$, the redshift will change by an
amount 
$\delta z$
\be
\delta z\equiv z(t_e + \delta t_e,t_0 + \delta t_0) - z(t_e,t_0)~.\lb{delz}
\ee
It is easy to derive the following expression
\bea
\delta z &=& \frac{\delta t_0}{a_e} \left( \dot{a}_0 - \dot{a}_e
\right)\lb{delz1} \\
&=& H_0 \delta t_0 \left( 1 + z(t_e,t_0) - h_e \right)~~~~~~~~~~h_e\equiv
\frac{H_e}{H_0}\\
&=& H_0 \delta t_0 \left( 1 + z - h(z) \right)\lb{delz2}
\eea 
with the obvious notation $H_e=H(t_e)$. We have dropped the subscript $e$ and 
finally we return to more conventional notations setting $z\equiv z(t_e,t_0)$. 
It is obvious from (\ref{delz1}) that $\delta z$ is a decreasing negative
function of $z$ 
in a universe whose expansion rate is always decelerated. This is what would
happen 
e.g. for an Einstein-de Sitter universe.   
However the situation changes when at least part of the expansion is
accelerated. 
Let us consider a universe for which the deceleration parameter $q$ satisfies
$q<0$ 
for $z<z_q$ and $q> 0$ for $z>z_q$ and $q(z_q)=0$. This is the case for a flat
universe 
with dust-like matter and a cosmological constant $\Lambda$ satisfying 
$\Omega_{\Lambda,0}>\frac{1}{3}$. 
For this universe we see from (\ref{delz1}) that $\delta z$ must be an
increasing 
function of $z$ on the interval $z<z_q$ during which the expansion rate is 
accelerating. At large redshifts, when the expansion rate is decelerating 
(and matter-dominated), $\delta z$ will decrease in function of $z$. 
Note that $\delta z$ vanishes in a Milne (empty) universe. 

It is interesting to study this behaviour by considering the slope 
$(\delta z)'\equiv \frac{d}{dz}\left( \delta z \right)$. 
We have 
\be
(\delta z)' = H_0 \delta t_0 \left(1 - h'\right)~.\lb{delz'}
\ee
The slope is positive for $h'<1$ and negative for $h'>1$. For a universe with 
accelerated expansion rate on the interval $0\le z \le z_q$, $\delta z$ 
reaches its maximum at $z_m$ when $h'(z_m)=1$. 
One can show that we have $z_m>z_q$ with $z_m$ typically only slightly larger
than $z_q$. 
We will consider below even more sophisticated scenarios where the slope is 
first negative at $z=0$, then positive at higher $z$ in order to ensure a
late-time 
accelerated stage and then again negative. 
It is convenient to introduce the new dimensionless quantity 
\be
\Delta z \equiv \frac{\delta z}{H_0 \delta t_0}~. 
\ee
It is easy to derive the following equalities
\bea
1 + q  &=& \frac{1-(\Delta z)'}{1-\frac{\Delta z}{1+z}}= \frac{3}{2} (1+w_{\rm
eff})\lb{q}\\
w_{DE} &=& \left[ \frac{1-(\Delta z)'}{\frac{3}{2} \left( 1-\frac{\Delta z}{1+z}
\right)} -1 \right]
    \left[1 -  \frac{\Omega_{m,0} (1+z)}{\left(1-\frac{\Delta z}{1+z}\right)^2}
\right]^{-1}~,\lb{w}
\eea
where spatial flatness is assumed in (\ref{w}). 
When we use luminosity distances, second derivatives are necessary.  
We have in particular 
\be
(\Delta z)'_0 = -q_0~.\lb{delz'0}
\ee 
In contrast to the luminosity distance $d_L(z)$, the leading order of 
$\delta z$ for $z\ll 1$ is model dependent and allows to discriminate between 
models with same $H_0$ but different $q_0$. 
Physically this is so because the slope at $z=0$ is sensitive to the 
dark energy EoS as we see from eqs.(\ref{q},\ref{w}). 
Note that at the redshift $z_1\ne 0$ where $\delta z=0$ (if it exists), 
$(\Delta z)'= - q$. 
The following inequality 
\be 
(\Delta z)' > 1~,\lb{ineq}
\ee
is satisfied by a universe on all redshifts for which it is in a 
phantom phase, $q<-1$ or $w_{\rm eff}=\Omega_{DE}~w_{DE}<-1$.
Hence a universe in a phantom phase on small redshifts must satisfy 
inequality (\ref{ineq}) on these redshifts. 
If it is DE itself which is of the phantom type, $w_{DE}<-1$, the following 
weaker inequality has to be satisfied 
\be
(\Delta z)' > 1 - \frac{3}{2} \frac{\Omega_{m,0} (1+z)^2}{1+z -\Delta z}~.
\ee
When DE is a cosmological constant $\Lambda$, the above inequality becomes a 
strict equality. Note that $(\Delta z)'= 1$ in a de Sitter universe which 
corresponds to the asymptotic future of a flat universe with a cosmological 
constant $\Lambda$ ($\Omega_m\to 0$).

These properties can be seen with the small-$z$ expansion 
\be
\Delta z = - \left[ q_0 z - \frac{1}{2} ( q^2_0 -j_0)z^2 +...\right]~,\lb{exp}
\ee
which is to be contrasted with the expansion 
\be
D_L(z)\equiv  H_0~d_L(z) =  z + \frac{1}{2} (1-q_0) z^2 +...\lb{DL} 
\ee
The leading term of (\ref{DL}) is the same for all models with identical $H_0$ 
in contrast to (\ref{exp}).
The result (\ref{delz'0}) is obviously recovered from (\ref{exp}). 
In (\ref{exp}) $j_0$ is the ``jerk'' parameter, $j=\frac{\dddot{a}}{aH^3}$, at
z=0.
Expression (\ref{exp}) shows also how $\Delta z$ deviates from a linear law at
small 
redshifts. From (\ref{exp}), this deviation is positive i.e. 
\be
\Delta z > - q_0 z~,~~~~~~~~~~~~~~~~~~~~~~z>0,~z\approx 0 \lb{exp1}
\ee
on some small interval around $z=0$ when the acceleration is slowing down 
on small redshifts. These properties of the redshift drift are illustrated 
with Figure 1.

Let us return finally to the amplitude of the redshift drift. Clearly it is 
proportional to the small quantity  $H_0 \delta t_0$. In other words all
equations 
and inequalities derived above make use of $\Delta z$, the redshift drift in
units 
$H_0 \delta t_0$. For $\delta t_0=N$ years, $H_0 \delta t_0=1.023~N~h \times
10^{-10}$ 
(here $h=H_0/100 km/s Mpc$), so $\delta t_0$ should be at least of the order of 
$10$ years to yield a measurable effect.
We will consider in the next section some specific models.

\section{Models with peculiar equation of state}

We consider now various DE models exhibiting a characteristic 
signature in their redshift drift because the equation of state 
(EoS) parameter $w_{DE}$ has a special behaviour on low redshifts.
The choice of models in this study looked for distinctive features. 

\begin{figure}[t]
\begin{centering}\includegraphics[scale=.9]{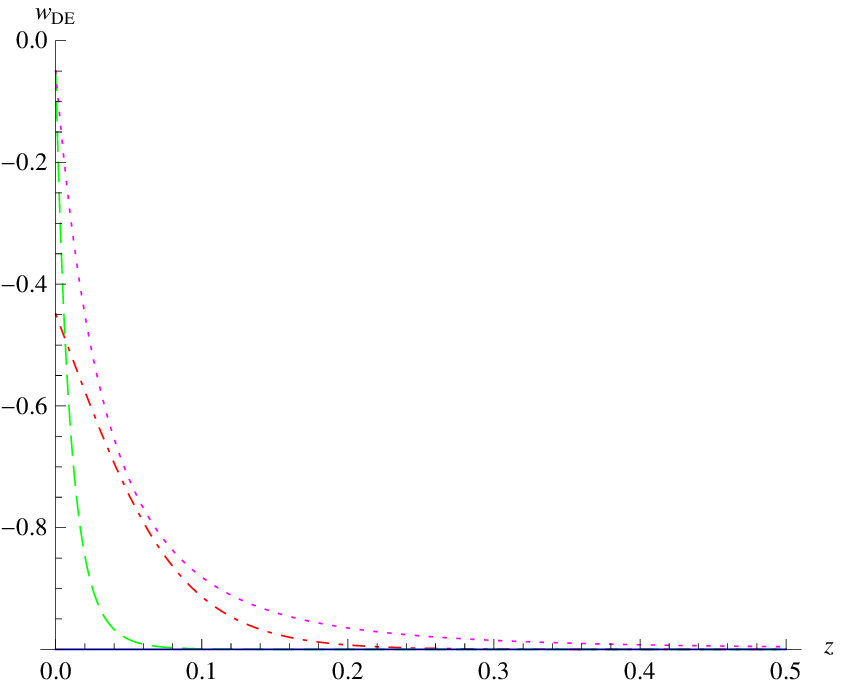}~~
\includegraphics[scale=.9]{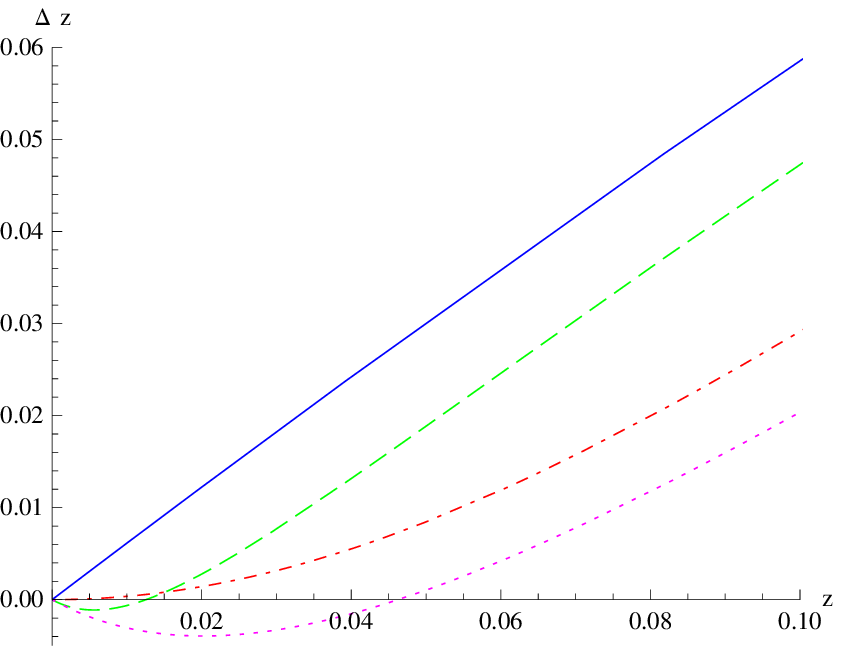}
\par\end{centering}
\caption{ a) On the left panel, three models are displayed with a 
large variation of $w_{DE}$ on low redshifts. In these universes the 
acceleration is slowing down on low redshifts and the expansion is even 
decelerating on very low redshifts for the top (pink) and bottom 
(green) curves. The red curve (in the middle) is taken from \cite{SSS09} 
and has $q_0\approx 0$. 
b) On the right panel the corresponding redshift drifts are displayed. The 
blue curve corresponds to a $\Lambda$CDM universe. Note that the other 
three curves satisfy the inequality (\ref{exp1}). 
The corresponding luminosity distances have no characteristic signatures 
though the pink curve is in tension with SNIa data while the model with the 
green curve provides an excellent fit because strong variation of 
$w_{DE}$ for this model, with presently decelerated expansion, is confined 
to the interval $z\le 0.1$.}
\lb{Fig1}
\end{figure}

\subsection{Large variation of $w_{DE}$ on low redshifts}
We start with models with a large variation of $w_{DE}$ starting from 
very low redshifts and on until today. There are several situations where 
such a variation can occur, for example in models where the accelerated 
stage is transient and the accelerated expansion is already slowing down 
on very low redshifts. This can be the case for some quintessence models 
(see e.g. \cite{BP04},\cite{DHRS09}) or for more exotic models with a 
singularity in the future (see e.g. \cite{KGGMK09}). 
It was also proposed recently as an interesting phenomenological ansatz that 
can improve the fit to SNIa data compared to a pure cosmological constant 
\cite{SSS09}. Three models are shown on Figure \ref{Fig1}.
We see that these models have a characteristic slope of their redfshift drift 
at very low redshifts which differs significantly from the slope obtained 
for $\Lambda$CDM $(\Delta z)' = 1 - \frac{3}{2} \frac{\Omega_{m,0} (1+z)^2}{1+z
-\Delta z}$.  
The slope at $z=0$ is negative if the universe is presently decelerating and
rapid variation of $w_{DE}$ produces a sharp change in $\delta z$.    
Redshift drift measurements on such very low redshifts suffer from increased 
difficulties related to the peculiar velocities of objects and are virtually 
impossible. While the signatures of the models considered in this subsection 
are most probably not observable, they illustrate nicely the results derived 
in section 2. As a substantial amount of SNIa data have been added in the 
interval $0.1\le z\le 0.4$ in the Union2 set as compared to the Constitution 
set, two of these models are now in tension with SNIa data. The model with the 
pink curve on Figure 1 (denoted mod2 in Table 1) is in tension with the data 
because it departs from $w_{DE}=-1$ on higher redshifts $z<0.5$.
However the model with the green curve in Figure 1 (denoted mod1 in Table 1) 
provides an excellent fit to the data. It has a sharp departure from $w_{DE}=-1$ 
on redshifts $z<0.1$ only and a decelerated expansion today. This is in agreement 
with results obtained in \cite{MHH09}.

\begin{figure}[t]
\begin{centering}\includegraphics[scale=.9]{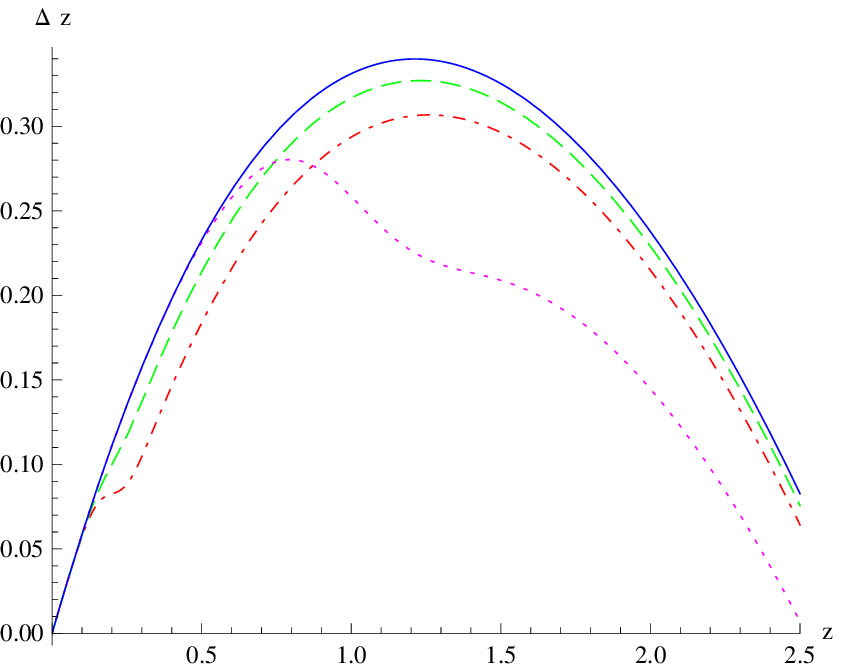}~~
\includegraphics[scale=.9]{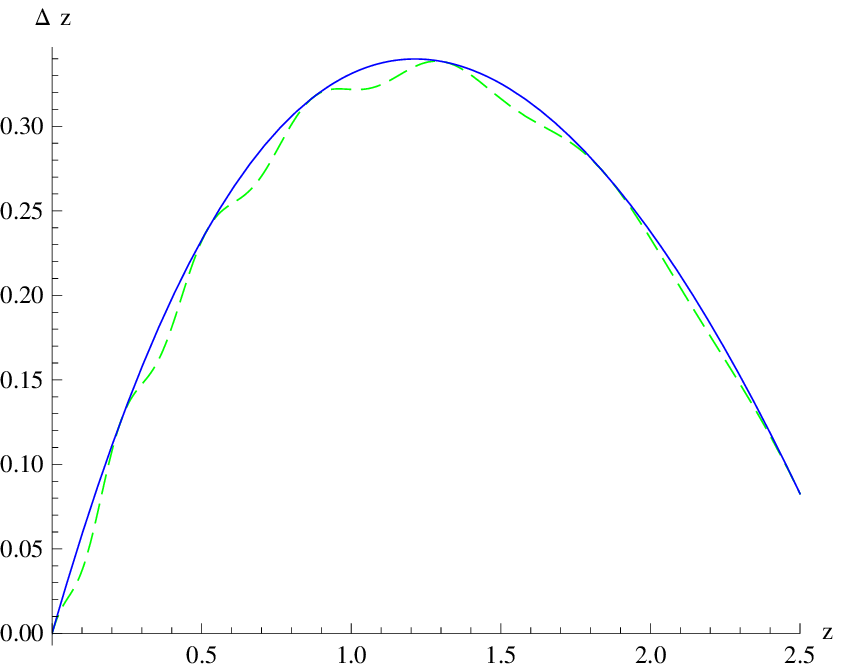}
\par\end{centering}
\caption{ a) The redshift drift for several models with a ``spike'' in 
their EoS parameter $w_{DE}$ is shown on the left panel. The
spikes located at $z=0.2$ (red and green curves) have 
identical widths and amplitudes (maximal heighths) $A_{sp}=0.5$ and $A_{sp}=0.2$ 
respectively. The third spike at $z=1$ (pink curve) has a larger width and 
amplitude $A_{sp}=0.7$. The location and amplitude of the spike leaves a neat 
signature in the redshift drift.    
b) The redshift drift of $\Lambda$CDM and a model with 
oscillating $w_{DE}$ (with amplitude $A_{osc}=0.3$) are displayed on the right 
panel. The oscillations are clearly seen in the redshift drift while 
they are essentially erased in the luminosity distance. } 
\lb{Fig2}
\end{figure}

\begin{figure}[t]
\begin{centering}\includegraphics[scale=.9]{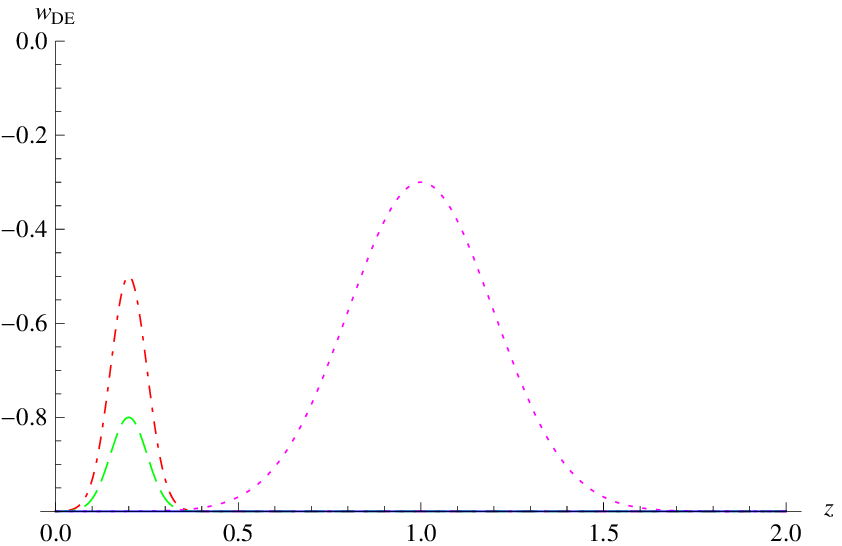}~~
\includegraphics[scale=.9]{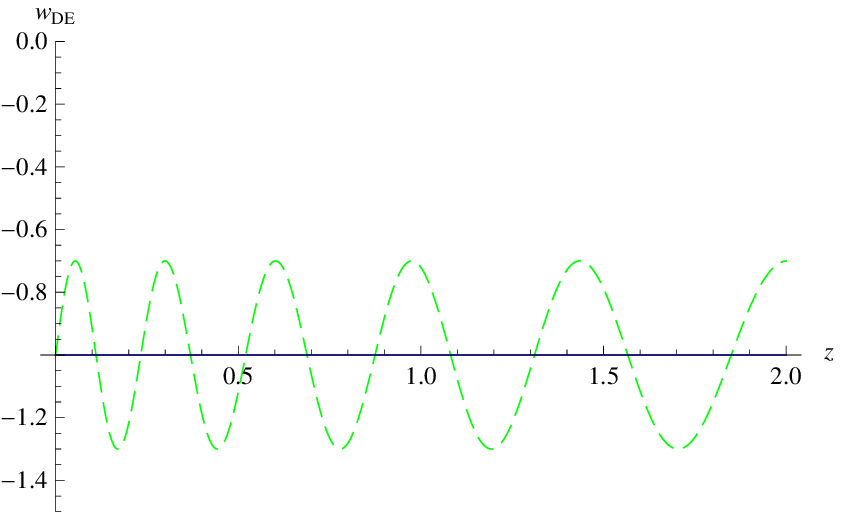}
\par\end{centering}
\caption{ a) The equation of state parameters of our models 
with spike appearing in Figures 2 (same colour as in Figure 2) 
are shown on the left panel. 
b) On the right panel, the oscillating model used in Figure 2 is 
displayed.} 
\lb{Fig3}
\end{figure}

\begin{figure}[t]
\begin{centering}\includegraphics[scale=.9]{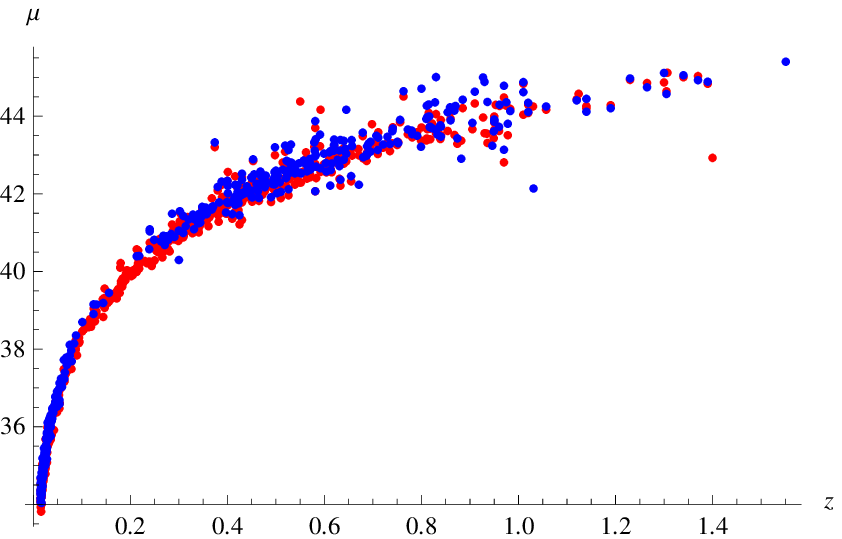}~~
\includegraphics[scale=.9]{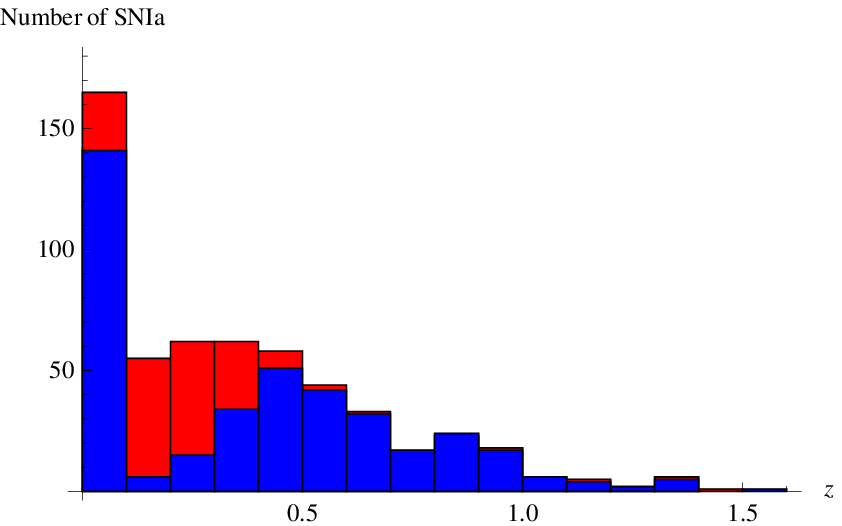}
\par\end{centering}
\caption{ a) SNIa distance modulus data from the Constitution set (blue points) 
and Union2 set (red points).    
b) Number of supernovae per redshift intervals of 0.1 for the two data sets.
There is a substantial addition in the Union2 data set in the interval
$0.1\le z\le 0.4$, which
accounts for the stronger constraints in models with modifications at low
redshift.}
\lb{Fig4}
\end{figure}

\subsection{Spikes in the equation of state}
Here we consider models where $w_{DE}$ exhibits a spiky feature (see e.g. 
\cite{MHH09},\cite{HPZ10}). The larger the redshift where a spike of given
amplitude (i.e. maximal heighth) $A_{sp}$, and width is located, the smaller the 
effect on $\delta z$ because $\Omega_{DE}$ decreases with redshift and 
hence the effect on $H(z)$ of a spike in $w_{DE}$ will decrease as well. 
We consider models which have the same fiducial value 
$w_{DE,0}=-1$ with a superimposed spike.
 
We can have models with a clear signature of the spike indicating precisely 
its location because the slope of $\delta z$ depends directly on $w_{DE}$ and 
several models are displayed in Figure \ref{Fig2}.
In luminosity-distances the feature of the spike is erased by the 
integration so the effect can easily be degenerate with variations of other 
cosmological parameters and there is no signature which can be unambiguously 
attributed to the presence of a spike in the equation of state parameter
$w_{DE}$.
The red curve with a pronounced spike at $z=0.2$ (denoted Spike R in Table 1) 
is ruled out by SNIa data while the green curve with a less pronounced spike 
located at the same redshift is in tension with the data. On the other hand 
the model with a spike at $z=1$ (denoted Spike P in Table 1) provides a very 
good fit to the data. It is interesting to remark that, due to
the reduced number of data at high redshifts, even a stronger modification in
$w_{DE}$ like this one can go unnoticed if only SNIa and BAO data is used. 
From Figure 2, the difference in the quantity $10^{10}~\delta z$ 
between the SpikeP model and $\Lambda$CDM at $z\sim 2$ in a 15-years experiment 
is estimated to be about 1.2. As things stand now this is still too small to 
allow for a discrimination by a planned experiment like CODEX \cite{Codex07}. 
However an improvement of order 4 of the sensitivity would allow to distinguish 
both models.  

\subsection{Oscillating dark energy}
Another possibility is provided by models where DE has an oscillating EoS
parameter 
(see e.g. \cite{L06}), 
viz. 
\be
w_{DE} = -1 + A_{osc} \sin \left[ B \ln (1+z) + C \right]~.\lb{wosc}
\ee
Oscillations in $H(z)$ are induced but will be essentially erased in the 
luminosity-distances. In contrast these oscillations appear clearly in the 
redshift drift as models displayed on Figure \ref{Fig2} show. 
It is quite clear that a detection of these oscillations require 
very accurate data. These oscillations could be detected with accurate data 
on redshifts $z\sim 2$ in the range where one expects at least in principle 
accurate data to be possible.  

\begin{table}
\begin{center}
\begin{tabular}{| c | c | c | c | c | c | c | c | c |}
\hline
& mod1 & mod2 & SSS & Spike R & Spike G & Spike P & Osc. & LCDM \\
\hline
$\Omega_{m,0}$ & 0.267 & 0.255 &
0.257 & 0.241 &
0.254 & 0.272 &
0.262 &
0.266\\
\hline
$\chi_2$ & 544.5 & 549.5 &
547.0 & 554.6 & 546.7 &
545.0 & 544.9 &
544.9\\
\hline
\end{tabular}
\end{center}
\caption{Best-fit $\Omega_{m,0}$ and $\chi^2$ values after 
marginalization of $H_0$, using the Union2 set \cite{Amanullah10} of 557 SN Ia,
the BAO distance ratio \cite{Percival10} and the CMB shift parameter \cite{K11}.
The first three models correspond resp. to the green, pink and red curves of 
Figure 1. The models Spike R, Spike G and Spike P correspond resp. to the red, 
green and pink curves of Figure 2a.
There are now enough SNIa data on low redshifts to put most of the models with a
large variation of $w_{DE}$, or a pronounced feature, on very low redshifts
either in tension with the data (Spike G, SSS) or even to rule out some of
them (Spike R, mod2). Redshift drift on these low redshifts would be virtually
impossible to measure. But other models are viable and could be potentially
constrained
using the redshift drift. In particular, Spike P is a strong
modification at large redshifts while fitting the data almost as well as
$\Lambda$CDM.} 
\label{Table1}
\end{table}

\section{Some models beyond General Relativity}

In this section we would like to show that redshift drift measurements could 
help 
in some cases to 
establish whether a model is inside General Relativity (GR) 
or outside it. 
We have in mind models outside GR for which the gravitational constant defining 
the Chandrasekhar mass evolves with time so that the intrinsic luminosity of SNIa 
would not be constant for such models. We caution that there are many modified gravity 
models where this is not the case because the gravitational constant around compact 
objects is constant and basically equal to the usual Newton's constant (see \cite{GMP08} 
for a related discussion in the context of some viable $f(R)$ DE models).  

For our discussion it is sufficient to consider spatially flat universes.
Let us assume that the intrinsic luminosity of SNIa has some redshift
dependence 
\be
L(z) = L_0 ~{\cal G}(z)~~~~~~~~~~~~~~~~~~~~~~~~{\cal G}(z=0)=1.\lb{Lz}
\ee
In principle this could also arise for physical reasons unrelated to 
modifications of gravity (see e.g.\cite{LVT09}), 
but we will have in mind DE models where the variation of $L$ is induced by 
the variation of the gravitational constant defining the Chandrasekhar mass. 
For example (\ref{Lz}) would occur in some models with varying gravitational 
coupling (see e.g. \cite{SSS05}) and in scalar-tensor models.  

From (\ref{Lz}), the expression for the measured flux ${\cal F}$ of 
SNIa can be written as 
\be
{\cal F} = \frac{L(z)}{4\pi d^2_L} = \frac{L_0}{4\pi {\cal D}^2_L}~,\lb{F}
\ee
where we have introduced the quantity 
\be
{\cal D}_L = d_L~{\cal G}^{-\frac{1}{2}}~.
\ee
Measuring the flux ${\cal F}$ does not allow to disentangle the expansion of 
the universe from the variation of the intrinsic luminosity. One can only 
recover from SNIa data the quantity ${\cal D}_L(z)$.

Confusing the two quantities $d_L$ and ${\cal D}_L$ would lead to 
an erroneous retrieval of the background expansion because  
$H^{-1}(z)\ne \left( \frac{{\cal D}_L(z)}{1+z}\right)'$.
We have instead the equality
\be
\left( \frac{{\cal D}_L}{1+z}\right)' = {\cal G}^{-\frac{1}{2}}~
   \left[ H^{-1} + \frac{d_L}{1+z} \left(\ln {\cal G}^{-\frac{1}{2}}\right)'
\right]~.\lb{tdL}
\ee
Measuring $H(z)$, and hence indirectly also $d_L(z)$, from the redshift drift 
allows us to check whether one has ${\cal G}\ne 1$ and also provides us with 
information about the behaviour of ${\cal G}(z)$.  

In the rest of this Section we consider the specific case of (massless) 
scalar-tensor models.   
In this model, the usual Newton's constant in the equation for the growth of 
perturbations (and in the Poisson equation) is replaced by the effective 
gravitational coupling constant $G_{\rm eff}(z)$, viz. \cite{BEPS00}
\be
h^2~\delta_m'' + \left(\frac{(h^2)'}{2} -
\frac{h^2}{1+z}\right)\delta_m' = \frac{3}{2} (1+z) \frac{G_{\rm eff}(z)}{G}~
\Omega_{m,0}~\delta_m~.\label{delm}
\ee
Knowing $H(z)$ and the perturbations $\delta_m(z)$, one can check whether 
these two functions are consistent within a given model.
The quantity $G_{\rm eff,0}\equiv G_{\rm eff}(z=0)$ is the coupling constant 
measured in a Cavendish type experiment, its numerical value is therefore 
extremely close to $G$. 

Taking into account the dependence of the Chandrasekhar mass on $G_{\rm eff}$
and assuming simple SNIa models, the peak luminosity turns out to be 
proportional to $(\frac{G_{\rm eff}}{G_{\rm{eff},0}})^{-\frac{3}{2}}$ and 
we have \cite{ACO99}
\be
{\cal G}(z) = \left(\frac{G_{\rm eff}}{G_{\rm{eff},0}}\right)^{-\frac{3}{2}}~,
\ee
leading to the well-known modification of the distance modulus $\mu$
\be
\mu = 5 \log {\cal D}_L + \mu_0 = 
  5 \log d_L + \frac{15}{4} \log \left(\frac{G_{\rm eff}}{G_{\rm{eff},0}}\right)
+ \mu_0~,\lb{mu}
\ee
with $\mu_0 = 25 + 5~\log~\left(\frac{cH_0^{-1}}{{\rm Mpc}}\right)$.
If the second term is unknown, we cannot retrieve $H(z)$ from $\mu$. On the
other hand if it is ignored, i.e. identifying ${\cal D}_L$ with $d_L$, an 
incorrect $H(z)$ is obtained from the observed $\mu$.
While ignoring this modification of $\mu$ may be enough to show at least an 
inconsistency with the growth of perturbations in $\Lambda$CDM, or more
generally in GR with smooth non-interacting DE, it does not allow to infer 
the correct behaviour of the quantity $G_{\rm eff}(z)$ from the growth of 
perturbations. 
On the other hand, redshift drift measurements yield directly the quantity
$H(z)$ 
\emph{independently} of SNIa data. Combining with SNIa data on a range of
redshifts 
where both data overlap, one can recover the behaviour of 
$G_{\rm eff}(z)$ in this interval applying (\ref{tdL}). The growth 
of matter perturbations give a consistency check with 
\be
h^2 = \frac{(1+z)^2}{\delta'^2}\Biggl[\delta_0'^2 + 3 \Omega_{m,0} \int_0^z 
                \frac{G_{\rm eff}}{G} \frac{\delta' \delta}{1+z'}
dz'\Biggr]~.\lb{h2RG}
\ee
Additional properties of the growth of matter perturbations can be used as
additional tests (see e.g. \cite{AV07}). 
Finally, let us mention that a varying $G_{\rm eff}$ will also affect the 
width of the light curve which could also provide some information about 
the evolution of $G_{\rm eff}$.  
\section{Summary and conclusions}

We have studied models for which accurate redshift drift data can exhibit 
distinctive features not present in luminosity-distances. 
We emphasize that another incentive to achieve accurate redshift drift 
measurements comes from 
some DE models outside GR where an evolving gravitational constant affects 
the intrinsic luminosity of SNIa. 
It is well-known that these models can be efficiently probed by checking 
whether the expansion rate $H(z)$ and the perturbations $\delta_m(z)$ are 
consistent with each other inside a given model. 
The redshift-drift gives a direct probe of $H(z)$ independently of the 
modifications of gravity in contrast to SNIa data and we have illustrated 
this with scalar-tensor DE models. 
The expansion rate on very low redshifts can also be probed with Baryon 
Acoustic Oscillations (BAO) (see e.g. \cite{BH09}) which rely on the nature 
and the evolution of matter perturbations. Systematics as well as the probed 
range of redshifts are different, so the two methods would be complementary. It
is interesting that models with a strong variation of their
EoS both on redshifts $z\le 0.1$ and $z\sim 1$ would probably escape all 
observations which leaves some uncertainty about the present and future 
evolution of our universe (see for example ourmod1 and Spike P). Modifications 
at high redshifts, in particular, are poorly constrained by current data but 
can be potentially probed by redshift drift measurements.

We caution that the effect is tiny and that it is not clear yet 
whether it could be measured with the required accuracy. 
This obviously represents a technological challenge. 
However, as the exquisite data of the Cosmic Microwave Background 
anisotropy have spectacularly shown (see e.g. \cite{K11}), technological 
progress 
could make such measurements possible.
If this is the case, redshift drift data could become useful  
as a complementary probe    
in order to help unveil the nature of DE especially if DE is 
of the kind investigated here. 

\section*{Acknowledgments}
BM thanks the financial support from the Research Council of Norway
(No.\,202629V11) during his stay at the ITA - Univ. Oslo. He also thanks the support of the
Laborat\'{o}rio Interinstitucional de e-Astronomia (LIneA) operated jointly by the
Centro Brasileiro de Pesquisas Fisicas (CBPF), the Laborat\'{o}rio Nacional de
Computa\c{c}\~{a}o Cient\'{i}fica (LNCC) and the Observat\'{o}rio Nacional (ON) and funded by
the Ministry of Science and Technology (MCT). 
  

\end{document}